\documentstyle[aps,prl,psfig,floats]{revtex}

\begin{document}

\draft \voffset 1cm

\twocolumn[\hsize\textwidth\columnwidth\hsize\csname %
@twocolumnfalse\endcsname

\title{Exponential Temperature Dependence of Penetration Depth in MgB$_2$}
\author{F.~Manzano and A.~Carrington}
\address{H.~H.~Wills Physics Laboratory, University of Bristol, Bristol, BS8 1TL, England.}
\date{\today}
\maketitle

\begin{abstract}
The temperature dependence of the London penetration depth, $\lambda(T)$, was measured in polycrystalline MgB$_2$
samples by a high-resolution, radio frequency technique. A clear exponential temperature dependence of $\lambda(T)$ was
observed at low temperature, indicating $s$-wave pairing. A fit to the data gives an effective energy gap $\Delta$ of
$2.8\pm0.4$ meV ($2\Delta/k_BT_c=1.7\pm0.2$), which is significantly smaller than the standard BCS weak coupling value
of 3.5. This may be explained either by an anisotropic gap or the influence of low frequency phonon modes.
\end{abstract}

 ]

The recent discovery \cite{nagamatsu} of superconductivity at 39~K in the binary compound MgB$_2$ has sparked a large
number of investigations into its physical properties.  A crucial question is whether its high $T_c$ can be explained
by a phonon mediated pairing interaction within the usual BCS-Eliashberg framework. A first step in answering this
question is to determine the symmetry of the superconducting order parameter and the nature of the low energy
excitations.

Several experiments so far, have given strong indications that the superconductivity is of the $s$-wave BCS type.  An
important experiment has been the observation of a large boron isotope effect \cite{Budko}. In conventional
superconductors, tunneling measurements played a key role in verifying the BCS theory, however, in the cuprate
superconductors this probe has not proved quite as useful because the short coherence length $\xi$ makes junction
fabrication difficult. The majority of tunneling measurements on MgB$_2$ so far show a $s$-wave like gap but the values
of the deduced gap vary widely between 2 and 7 meV \cite{Giubileo,ctchen,karapetrov,Bollinger}. The magnetic
penetration depth $\lambda$ has proven to be a powerful probe of the gap symmetry in {\it both} classical and high
$T_c$ superconductors and gives unambiguous information about the nature of the low energy excitations in
superconductors. As the length scale probed is of order $\lambda$ rather than $\xi$, for high $\kappa$ materials,
penetration depth measurements are much less sensitive to surface quality than tunneling, and hence we expect the
results to be more representative of the bulk.

There have been several recent reports of measurements of $\lambda(T)$ for  MgB$_2$  by ac-susceptibility
\cite{panagopoulos,chen}, muon spin rotation \cite{panagopoulos} and optical conductivity \cite{pronin} techniques.
These authors conclude that at high temperature  $\lambda(T)$ follows a power law dependence [$\lambda(T)\sim T^2$
(Refs.\ \cite{panagopoulos,pronin}) and $\lambda(T)\sim T^{2.7}$ (Ref.\ \cite{chen})], which seems to be at odds with
tunneling and other measurements which indicate that there is a sizeable $s$-wave gap.  In this paper we present high
resolution measurements of the temperature dependence of the penetration depth in polycrystalline samples of MgB$_2$.
We find strong evidence for a predominately exponential temperature dependence of $\lambda$ at low temperature
consistent with $s$-wave BCS behavior. The gap deduced from fits to the data however, is significantly smaller than the
BCS weak coupling value.  This may be explained by either a strong anisotropy of the gap or the presence of low energy
phonon modes.

Three different types of polycrystalline MgB$_2$ samples were studied.  Samples A and C, were produced from
commercially available (Alfa Aesar) powder.  Sample A was produced by casting the powder directly in epoxy resin
(weight ratio MgB$_2$:Epoxy $\sim$ 1:6.5).  The commercial powder has a relatively wide range of grain sizes which
complicates quantitative analysis of the data.  For the analysis to give a reliable estimate of the absolute values of
$\lambda(T)$ grain sizes $\lesssim 5 \lambda$ are required \cite{porch93}.  To obtain such a distribution the
commercial powder was ground in an agate mortar and then sedimented in acetone for 1 hour\cite{athanassopoulou}.  The
resulting powder was then cast in epoxy ($\sim$6\% by volume) to produce sample C. The grain size distribution was
measured from scanning electron microscope images. The diameter of 96\% of the grains was less than 1$\mu$m, and the
mean diameter was 0.56$\mu$m. Sample B was fabricated by reacting boron powder and Mg flakes in an Argon atmosphere at
high temperature. This procedure produced a dense polycrystalline pellet. X-ray diffraction and scanning electron
microscopy showed that as grown samples of type B contained some residual flakes of unreacted Mg. Initial measurements
found that this changed the low temperature behavior of $\lambda(T)$ markedly \cite{ruslan}. To remove this
contamination we immersed these samples in a dilute solution ($\sim$ 0.5 \%) of HCL and ethanol. $T_c$ was unaffected
by this. Previous studies have shown that the commercial MgB$_2$ powder is 98\% phase pure, the remaining 2\% being
mostly composed of iron, carbon and oxygen \cite{wang}.  The susceptibility of these impurities is too weak to effect
the results presented here \cite{wang,cooper96}

Measurements of penetration depth were performed in a tunnel diode oscillator operating at 11.9~MHz\cite{carrington99},
with frequency stability of a few parts in $10^{10}$~Hz$^{-\frac{1}{2}}$.  For the powder samples in the present work
this translates to a resolution in $\lambda$ of $10^{-12}$m. A particular feature of our apparatus is the very low
value of the ac-probe field which we estimate to be $\sim 1 \mu$T. Ambient dc fields are shielded to a similar level
with a mu-metal can.  Changes in the oscillator frequency are directly proportional to the inductance of the probe coil
and hence to the susceptibility of the sample.  For single crystal samples, this frequency shift may be directly
related to changes in the penetration depth using the known sample dimensions \cite{prozorov00}. In general for
polycrystalline samples the relation between the measured susceptibility (per unit volume of superconductor) $\chi$ and
$\lambda$ is more complicated and depends on the size distribution of the grains.  For well separated grains
\begin{equation}
\chi = \frac{-\frac{3}{2}\sum\limits_i \left(1-
\frac{3\lambda}{r_i}\coth(\frac{r_i}{\lambda})+\frac{3\lambda^2}{r_i^2}\right)r_i^3 N_i }{\sum\limits_i r_i^3 N_i }
\label{chieq}
\end{equation}
where $N_i$ is the measured number of grains of radius $r_i$ \cite{porch93}.  For sample C, $\lambda(T)$ was determined
from the measured $\chi(T)$ and grain size distribution by solving Eq.\ (\ref{chieq}) at each point.  For samples A and
B the wide distribution of grain sizes and (for sample B) the possibility of interconnection between the grains makes a
quantitative determination of $\lambda(T)$ impossible.  We note however, that if $\lambda(T)$ becomes temperature
independent, then so will $\chi(T)$, irrespective of the detailed dependence of $\chi(T)$ on $\lambda$ and $r_i$.

\begin{figure}
\centerline{\psfig{figure=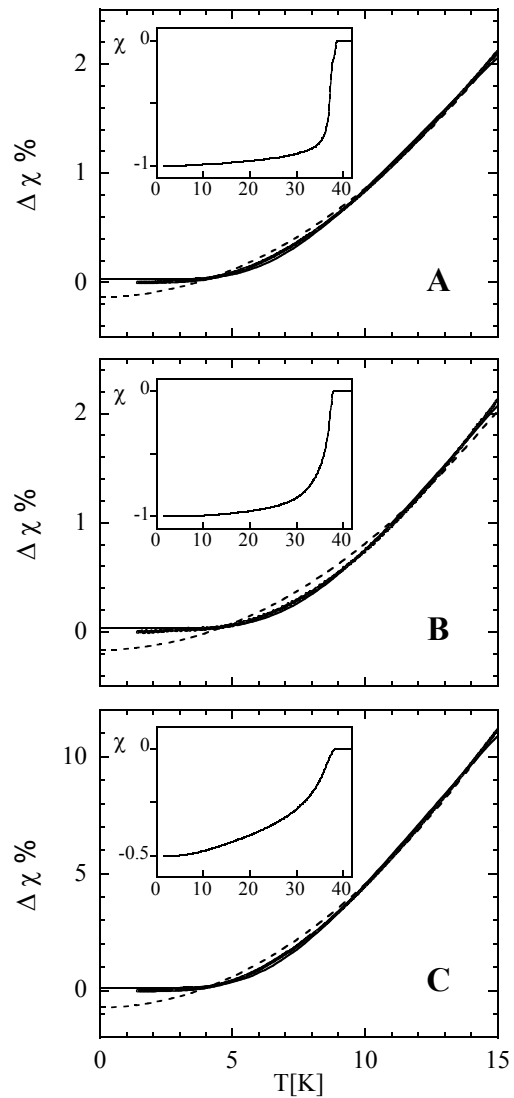,width=7cm}} \vskip 0cm \caption{Temperature dependence of the susceptibility
($\chi$) of samples A, B and C.  At low temperature (main panels) $\Delta \chi$ is directly proportional to the
temperature dependent penetration depth $\Delta \lambda(T)$.  The solid line is a fit to the $s$-wave BCS form [Eq.\
(\protect\ref{bcsfit})], and the dashed line to a $T^2$ behavior.  The insets show $\chi(T)$ over the full temperature
range.} \label{chifig}
\end{figure}

In Fig.\ \ref{chifig} we show the temperature dependence of $\chi$ for the three different samples. For samples A and
B, $\chi$ was normalized to $-1$ at low temperature, whereas for sample C,  $\chi$ was calculated from the measured
volume of superconductor and is thus in absolute units.  $\Delta \chi(T)$ denotes the change in $\chi$ relative to its
value at our base temperature ($T$=1.4~K). It is apparent that in all three samples $\Delta \chi(T)$ becomes
essentially temperature independent below around 5~K. This is the main result of this paper and is a clear indication
of an finite energy gap in all directions in $k$-space. A large distribution of grain sizes or anisotropy of $\lambda$
will effect the detailed $T$ dependence of $\chi$ (especially at higher temperature) but cannot cause such an abrupt
flattening off of $\Delta \chi(T)$.

The material parameters of MgB$_2$ deduced from other measurements put it firmly in the clean, local limit
($\lambda(0)\sim 0.2 \mu$m , $\xi(0)\sim $50 \AA~ and mean free path $\ell\sim$ 600 \AA \cite{finnemore,canfield}), so
$\Delta\chi(T)$ is directly proportional to changes in the London penetration depth ($\lambda_L$). For a $s$-wave
superconductor, at sufficiently low temperature, irrespective of the coupling strength, $\lambda_L(T)$ should follow
the BCS behavior
\begin{equation}
 \Delta \lambda(T) \simeq \lambda^e_0 \sqrt {\frac{\pi \Delta_0}{2k_BT}} \exp \left( \frac{-\Delta_0} {k_BT}\right)
\label{bcsfit}
\end{equation}
were $\Delta_0$ is the value of the energy gap at zero temperature and $\lambda^e_0=\lambda(0)$ in the weak coupling
case. In Fig.\ \ref{chifig} we show a fit of Eq.\ (\ref{bcsfit}) to the data for the three samples.  The fits give
consistent values for $\Delta_0$ which average $33\pm4$~K (the uncertainty reflects the different values of $\Delta_0$
between the three samples and as the upper limit of the fit ($T^{\text{fit}}_{\text{max}}$) was varied from 15~K to
20~K). It should be noted that Eq.\ (\ref{bcsfit}) is not a perfect fit to the data and particularly at very low
temperature where the measured $\Delta\lambda(T)$ has a larger temperature dependence that would be indicated by Eq.\
(\ref{bcsfit}).  If taken literally this may indicate the presence of a small amount of excitations with energies
significantly below our deduced $\Delta_0$. However, until high quality single crystals become available it is
impossible to rule out extrinsic origins of this term (for instance from weak links at the surface of the grains).  The
fit improves if two gaps are used (for instance with  $\Delta_1\sim 2\Delta_2$ and the lower gap having a weight of 5\%
of the larger gap), but this leaves the value of the larger gap unchanged (within the error) from our original
estimate.

As mentioned above, several other authors \cite{chen,panagopoulos,pronin} have claimed that $\lambda(T)$ in MgB$_2$
does not follow a simple exponential $T$ dependence but rather a power law dependence with an exponent close to 2. In
Fig.\ \ref{chifig} we show a $T^2$ fit to data. Clearly, this fit is much worse than the BCS dependence [Eq.\
(\ref{bcsfit})].   We have also tried to fit other forms such as $AT+BT^2$ or $AT^2/(T+T^*)$ but both fit the data much
worse than Eq.\ (\ref{bcsfit}) (the latter of these two equations is the form followed by a dirty $d$-wave
superconductor \cite{hirschfeld93}).  The key differences between our data and previous studies is our very high
resolution, low base temperatures and very low excitation fields.

\begin{figure}[t]
\centerline{\psfig{figure=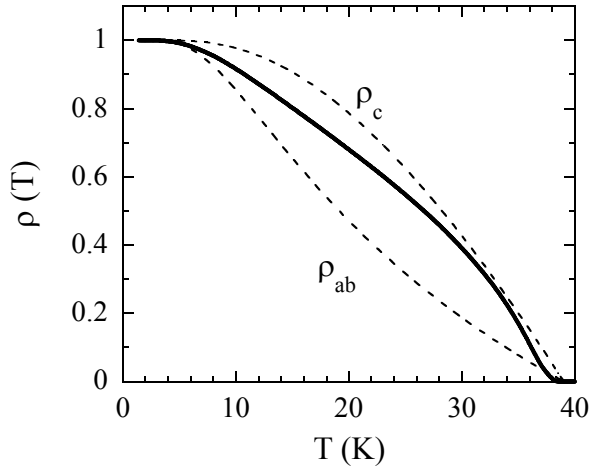,width=8cm}} \vskip 0cm \caption{Temperature dependence of the superfluid density
$\rho(T)=\lambda(0)^2/\lambda(T)^2$ for sample C (solid line). The dashed lines are the predictions of the anisotropic
gap model (see text)} \label{rhofig}
\end{figure}

\begin{figure}[t]
\centerline{\psfig{figure=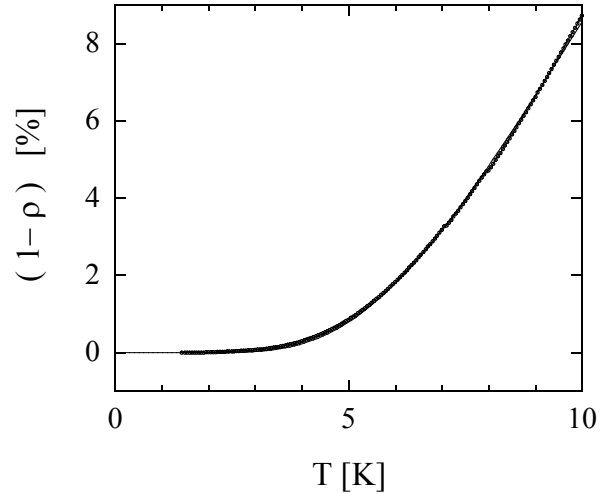,width=8cm}} \vskip 0cm
\caption{ Low temperature behavior of $\rho(T)$ for sample C along with a BCS fit (Eq.\ (\protect\ref{bcsfit})).}
\label{rhofig2}
\end{figure}

As described above, for sample C we are able to deduce the absolute values of $\lambda(T)$.  In Fig.\ \ref{rhofig} we
show the temperature dependence of the calculated superfluid density $[\rho=\lambda(0)/\lambda(T)]^2$.  Our data give a
value of $\lambda(0)=160\pm20$nm which is in good agreement with other estimates \cite{finnemore,chen} although is
somewhat longer than that deduced from $\mu$SR data \cite{panagopoulos}. The uncertainty in $\lambda(0)$ does not make
any significant difference to the temperature dependence of $\rho(T)$. The fit to the low temperature behavior of
$\rho(T)$ (Fig.\ \ref{rhofig2}) gives values of $\Delta_0$ in the range 28~K to 33~K as $T^{\text{fit}}_{\text{max}}$
was varied from 10~K to 15~K, in good agreement with those extracted from the fits to $\Delta\chi(T)$ in Fig.\
\ref{chifig}. From this data we are also able to calculate the constant $\lambda^e_0$ in Eq.\ (\ref{bcsfit}).  We find
that, unlike $\Delta_0$, $\lambda^e_0$  depends quite strongly on the fitting range - increasing from 57~nm to 90~nm as
$T^{\text{fit}}_{\text{max}}$ was varies from 10~K to 15~K.  This is however, significantly below the BCS value of
$\lambda(0)$= 160 nm.

MgB$_2$ has a anisotropic structure, with Mg atoms sandwiched between planar, hexagonal boron rings.  It is therefore
expected that there will be some anisotropy between the $ab$-plane and the $c$-axis responses. As our measurements have
been performed on unaligned polycrystalline samples our measurements of $\lambda(T)$ will be a mixture of the
$ab$-plane and $c$-axis penetration depths ($\lambda_c$ and $\lambda_{ab}$).  Recent results on aligned crystallites
and single crystals have shown a significant anisotropy in $H_{c2}$ which implies an anisotropy in the coherence
lengths, $\gamma=\xi_{ab}/\xi_{c}$.  In an anisotropic Ginzburg-Landau theory this implies a similar anisotropy in
$\lambda$,  $\gamma=\lambda_c/\lambda_{ab}$. There is some disagreement about the magnitude of this anistropy, Refs.\
\cite{lima,jung} give $\gamma=1.7\pm0.1$ and Refs.\ \cite{xu,lee} give $\gamma=2.6\pm0.1$. As far as we know the is no
general solution for the moment of a sphere when $\lambda$ is anisotropic, however solutions do exist in the limits
$\lambda \gg r$ \cite{kogan} and $\lambda \ll r$ \cite{kufaev}. Fortunately, these two limits give similar results for
$\gamma\lesssim 3$. To within $\pm$10\% the effective $\lambda(0)$ equals 1.2  $\lambda_{ab}$ or 1.5 $\lambda_{ab}$ for
$\gamma$=1.6 and 2.6 respectively. Our value of $\lambda(0)$ then implies that $\lambda_{ab}$=130nm and
$\lambda_c$=210nm or $\lambda_{ab}\simeq 110$ nm  and $\lambda_c\simeq 280$ nm, for the two values of $\gamma$
respectively.

Although the overall temperature dependence of $\lambda$ deduced from our measurements is consistent with the a
$s$-wave BCS picture, the value of the gap obtained is somewhat smaller than the usual BCS weak coupling value
($\Delta/k_BT_c= 1.76$).  As pointed out by Haas and Maki \cite{haas} a simple explanation for this may be that the
$s$-wave gap is anisotropic.  In most conventional superconductors the anisotropy in the pairing interaction is weak
($\sim$ 10\%). However, it has been proposed that the unique structure of MgB$_2$ may produce a significantly enhanced
anisotropy \cite{voelker}.

Following Ref.\ \cite{haas} we model the anisotropy in $\Delta$ by $\Delta(z)=\Delta \Gamma(z)$ where $z=\cos\theta$
($\theta$ is the polar angle - with respect to the $c$ direction), $\Gamma(z)=(1+a z^2)/(1+a)$ and $a$ is the parameter
which controls the anisotropy. To determine $\Delta \lambda(T)$ in this model it is necessary to first calculate
$\Delta(T)$ self consistently from the gap equation
\begin{eqnarray}
\lefteqn{\frac{1}{N_0V}\int_0^1 \Gamma(z)^2 dz =} \nonumber \\ &&\int_0^1 \int_0^{\hbar \omega_c} \Gamma(z)^2 \frac{\tanh
\left(\frac{1}{2T}\left(E^2+\Delta(z)^2\right)^\frac{1}{2}\right)}{\left(E^2+\Delta(z)^2\right)^\frac{1}{2}}dE dz
\end{eqnarray}
and then $\Delta \lambda_{ab}(T)$ from
\begin{equation}
\frac{\Delta \lambda_{ab}^2(T)}{\lambda^2(0)} = 6 \int_0^1\!\!\int_{\Delta(z)}^\infty \frac{df(E)}{dE}
\frac{E}{\left(E^2-\Delta(z)^2\right)^\frac{1}{2}} z^2 dz dE
\end{equation}
and similarly for $\lambda_{c}(T)$ ( $f$ is the fermi function).  The temperature dependence of  $\Delta\lambda(T)$
measured on unaligned polycrystalline samples can then be calculated by a suitable average of $\Delta\lambda_{ab}$ and
$\Delta\lambda_{c}$. For moderate to high anisotropy the low temperature behavior is dominated by $\Delta\lambda_{ab}$
as in this model the gap is smallest in the plane.   The resulting $T$ dependence of $\lambda$ may then be fitted by
Eq.\ (\ref{bcsfit}) with an effective gap which is approximately equal to the minimum gap in the model (the in-plane
gap). We find that in order to explain the experimentally determined value of $\Delta/T_c=0.86\pm0.1$ we need
$a=2.2\pm0.4$, or a total gap anisotropy $(1+a) = 3.2\pm0.4$.  Whether or not this very strong anisotropy may be
explained by a phonon interaction alone needs to be addressed by microscopic theories. This anisotropy implies a
significant difference in the temperature dependence of both $\lambda(T)$ and $\rho(T)$ between the in-plane and
out-of-plane directions. The different $T$ dependencies of $\rho_c(T)$ and $\rho_{ab}(T)$ calculated from this model
for $a=3$ are shown in Fig.\ \ref{rhofig}.  It should be noted although the detailed behavior will be different if
strong coupling corrections and fermi velocity anisotropy are included, the qualitative difference between $\rho_c(T)$
and $\rho_{ab}(T)$ will remain. As our measurements were made on unaligned polycrystals, our data will be an average of
$\rho_c(T)$ and $\rho_{ab}(T)$ (the weights depending on anisotropy in $\lambda(0)$). Only a qualitative comparison is
possible, but it can be seen that the measured $\rho (T)$ falls between $\rho_c(T)$ and $\rho_{ab}(T)$.

A different approach has been taken by Manske {\it et al.} \cite{manske}, who have calculated the tunneling density of
states from first principles within a standard Eliashberg theory.  Their calculations suggest that the small value of
$\Delta_{min}/T_c$ is due to a pronounced low frequency dependence of $\Delta(\omega)$ originating from a low-frequency
phonon mode. The size of the minimum gap they calculate is in broad agreement with tunneling and our data.

In conclusion we have presented a study of the penetration depth of polycrystalline MgB$_2$ samples.  In agreement with
other direct probes of the symmetry of the order parameter we find that $\lambda(T)$ is well described by the usual
$s$-wave BCS behavior, but with a gap that is significantly below the weak coupling BCS value.  If this is to be
ascribed to a simple anisotropy of the gap then we find that a fairly large anisotropy of around 3 is required.
Alternatively the recent calculation of Manske {\it et al.} \cite{manske}  may also explain our data without the need
for significant anisotropy in $\Delta(k)$. Further studies on single crystal samples or aligned polycrystallites will
help to clarify this question.

We thank R.W.\ Giannetta, R.\ Prozorov, K.\ Maki, N.E.\ Hussey, J.F.\ Annett and J.R.\ Cooper for useful discussions,
and P.\ Timms for providing the type B MgB$_2$ samples.

\end{document}